\documentclass[lineno]{jfm}

\usepackage{graphicx}
\usepackage{newtxtext}
\usepackage{newtxmath}
\usepackage{natbib}
\usepackage{hyperref}
\hypersetup{
    colorlinks = true,
    urlcolor   = blue,
    citecolor  = black,
}

\newcommand{\RomanNumeralCaps}[1]
\linenumbers

\usepackage{graphicx}
\usepackage{natbib}

\usepackage{color,floatflt}
\usepackage{booktabs,subfigure}
\usepackage{amsfonts}
\usepackage{amssymb}

\usepackage{verbatim}
\usepackage{comment}
\usepackage{tikz}
\usepackage{tabularx}
\usepackage{bigstrut}
\usepackage{enumitem}

\usepackage{longtable}
\usepackage{comment}
\usepackage{soul}

\usepackage[bordercolor=black,linecolor=black,backgroundcolor=white,textwidth=3.5cm]{todonotes}
\usepackage{epstopdf}
\usepackage{lscape}

\makeatletter
\tikzset{
  RL/.style={
    rounded corners={1pt},
    to path={
      \pgfextra
        \pgf@process{\pgfpointanchor{\tikztostart}{east}}%
        \pgf@xa\pgf@x\pgf@ya\pgf@y
        \pgf@process{\pgfpointanchor{\tikztotarget}{east}}%
        \pgf@xb\pgf@x\pgf@yb\pgf@y
        \ifdim\pgf@xb>\pgf@xa
          \pgf@xa\pgf@xb
        \fi
        \pgfmathsetlength\pgf@xc{#1}%
        \advance\pgf@xa\pgf@xc
      \endpgfextra
      -- (+\pgf@xa,+\pgf@ya) -- (+\pgf@xa,+\pgf@yb) \tikztonodes -- (\tikztotarget)
    }
  },
  RL/.default=2cm}

\tikzset{
  LL/.style={
    rounded corners={1pt},
    to path={
      \pgfextra
        \pgf@process{\pgfpointanchor{\tikztostart}{west}}%
        \pgf@xa\pgf@x\pgf@ya\pgf@y
        \pgf@process{\pgfpointanchor{\tikztotarget}{west}}%
        \pgf@xb\pgf@x\pgf@yb\pgf@y
        \ifdim\pgf@xb>\pgf@xa
          \pgf@xa\pgf@xb
        \fi
        \pgfmathsetlength\pgf@xc{#1}%
        \advance\pgf@xa\pgf@xc
      \endpgfextra
      -- (+\pgf@xa,+\pgf@ya) -- (+\pgf@xa,+\pgf@yb) \tikztonodes -- (\tikztotarget)
    }
  },
  LL/.default=-2cm}

\ifCUPmtlplainloaded \else
  \checkfont{eurm10}
  \iffontfound
    \IfFileExists{upmath.sty}
      {\typeout{^^JFound AMS Euler Roman fonts on the system,
                   using the 'upmath' package.^^J}%
       \usepackage{upmath}}
      {\typeout{^^JFound AMS Euler Roman fonts on the system, but you
                   dont seem to have the}%
       \typeout{'upmath' package installed. JFM.cls can take advantage
                 of these fonts,^^Jif you use 'upmath' package.^^J}%
      }
  \else
  \fi
\fi

\ifCUPmtlplainloaded \else
  \checkfont{msam10}
  \iffontfound
    \IfFileExists{amssymb.sty}
      {\typeout{^^JFound AMS Symbol fonts on the system, using the
                'amssymb' package.^^J}%
       \usepackage{amssymb}%

      }{}
  \fi
\fi

\ifCUPmtlplainloaded \else
  \IfFileExists{amsbsy.sty}
    {\typeout{^^JFound the 'amsbsy' package on the system, using it.^^J}%
     \usepackage{amsbsy}}
    {\providecommand\boldsymbol[1]{\mbox{\boldmath $##1$}}}
\fi

\def\bfg#1{\setbox0=\hbox{$#1$}
        \kern-.025em\copy0\kern-\wd0
        \kern.05em\copy0\kern-\wd0
        \kern-.025em\raise.0433em\box0}

\def\drwln#1#2{\raise 2.5pt\vbox{\hrule width #1pt height #2pt}}

\newcommand{\dpr}[2] {\frac{\partial{#1}}{\partial {#2}}} 

\definecolor{darkgreen}{RGB}{0,128,0}

\graphicspath{{./figs/}}

\title{Thrust Generation by Shark Denticles}

\author{Benjamin S. Savino\aff{1} and Wen Wu\aff{1}
  \corresp{\email{wu@olemiss.edu}}}

\affiliation{\aff{1}Department of Mechanical Engineering, University of Mississippi, Oxford, MS 38677, USA
}

\begin{document}
\maketitle

\begin{abstract}
DNS is performed for flow separation over a bump in a turbulent channel. Comparisons are made between a smooth bump configuration and one where the lee side is covered with replicas of complete shark denticles. 
As flow over the bump is under an adverse pressure gradient (APG), a reversed pore flow (RPF) is formed in the porous cavities underneath the crowns of the denticle array. Remarkable thrust is generated by the RPF as denticle necks accelerate the fluid passing between them in the upstream streamwise direction. Several geometrical features of shark denticle, including some that had not previously been considered hydrodynamically functional, are identified to form an anisotropic permeable porous media that enables and sustains the RPF and thrust generation. The RPF is activated by the APG before massive flow reversal. The results indicate a proactive, on-demand drag reduction mechanism that leverages and transforms the APG into a favorable outcome. 
\end{abstract}



\section{Introduction}
Shark dermal scales exhibit notable distinctions from those found in bony fish. They are tiny (0.2-0.5mm in size) and display considerable variation in
shape and size~\citep{Reif85}. These unique structures, known as shark denticles, are believed to have the ability to reduce drag, consequently enhancing cruising efficiency and burst speed when hunting prey. The exploration of denticles' hydrodynamic functions can be traced back to the early 1980s~\citep{Reif85,BechertHR85} and has progressed in tandem with efforts to visualize, characterize, and control turbulent wall-bounded flows. While the streamwise-oriented ridges on their crown resemble the recognized drag-reducing riblets, researchers have not agreed upon the mechanisms of the hydrodynamic advantages of shark skin. Experimental tests with real shark skin rarely show drag reduction~\citep{BechertHR85,OeffnerL12}. Similarly, recent direct numerical simulations of turbulent channel flows over shark denticle replicas reported an increase in drag~\citep{BoomsmaF16,Lloydetal23}. 

Meanwhile, there has been a growing interest in investigating the potential of shark denticles to mitigate pressure drag by influencing flow separation.
 A series of studies by Lang and collaborators~\citep{Lang_2008,Langetal11,Santosetal21} focused on the denticle's ability to bristle, hypothesizing that bristling activated by reversed flow enhances turbulent transport of momentum.
\cite{Evansetal17} and \cite{Doosttalabetal18} simplified the denticles as diverging pillars and placed an array of them on the lee sides of an expanding channel and an airfoil. They observed a delayed separation and a reduced separation region. It was hypothesized that the forward pore flow between the pillars creates local unsteady suction and blowing. However, the lack of data in the immediate vicinity of the denticles hinders conclusive proof of these hypotheses.

While the mechanisms underlying the modulation of separation by shark denticles
remain unclear, studies have actively explored the impact of these structures on aerodynamic performance in various engineering applications. Denticle crown arrays~\citep{WenWL14,Domeletal18_low,Guoetal21}, real shark skin~\citep{OeffnerL12,Santosetal21}, and sparsely distributed denticle replicas used as vortex generators~\citep{Arunvinthanetal21,Chenetal23} have been applied to various separating flows. Diverse or contrasting results in the lift and/or drag have been observed. 
These studies often lacked explanation for the observed changes in performance,
obtaining detailed flow measurements around and below the denticles. 
In this study, we employ direct numerical simulation (DNS) to investigate the role of denticles during flow separation. In particular, we consider the complete denticle geometry including the neck. The simulation produces high-fidelity data on the local production of drag and flow fields at the sub-denticle scales. Correlations between the hydrodynamic performance indicators and the flow statistics are used to understand the physical mechanisms of drag modulation by sharks denticles.

\section{Methodology}
\begin{figure}
  \includegraphics[width=0.998\linewidth]{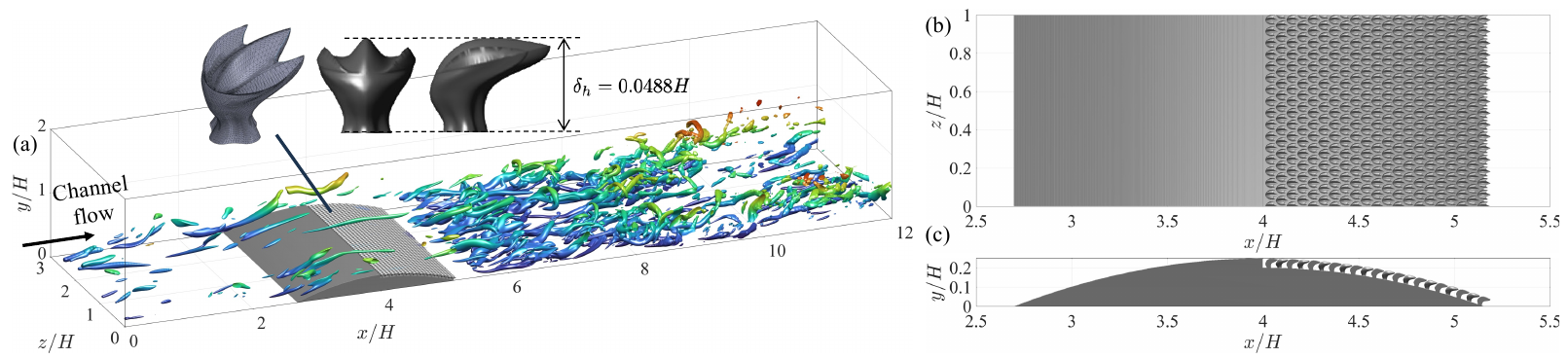}
  \caption{(a) Computational configuration. 
  An instantaneous flow field of case DT is shown, visualized by the isosurfaces of the second invariant of the velocity-gradient tensor
  and colored by the distance from the bottom wall. Inset, 3D model of the shark denticle~\citep{WenWL14}.  (b, c) Top and side views of the bump and denticle array.} 
  \label{fig:domain}
\end{figure}

We perform DNS of turbulent channel flows with a parabolic bump placed on the bottom wall to induce flow separation. This configuration has been utilized in our previous studies of flow separation~\citep{WuPS22} thus the baseline dynamics are well understood. The simulation is performed at $Re_b = U_bH/\nu = 2,500$. Here, $H$ is the channel half height and $U_b$ is the bulk velocity. These velocity and length scales are used to non-dimensionalize the incompressible Navier-Stokes equation and quantities discussed hereafter. Without the bump and denticles, the channel flow yields a friction Reynolds number $Re_\tau = u_\tau H/\nu = 160$ ($u_\tau$ is the friction velocity). 
The bump is defined by $y=-a(x-4.0)^2 + h$, where $x$ and $y$ are streamwise and wall-normal coordinates. The two parameters are $a$ = 0.15 and the bump height, $h=0.25H$ (figure \ref{fig:domain}(a)). 

Two simulations are performed and compared to justify the roles of denticles. The smooth bump is used in case SM. The lee side of the bump is covered by staggered array of stationary shark denticles in case DT. The decision to only cover the lee side of the bump is because the adverse pressure gradient (APG) here is more relevant to separating flows than the favorable pressure gradient (FPG) over the wind side, and a significant reduction in computation cost is achieved by covering only the former.
What differentiates our denticle array from previous studies are: first, the denticles do not protrude into the flow. In case DT, the lee side of the bump is indented by the height of the denticles. This ensures that the crests of the denticles align precisely with the contour of the smooth bump, rather than protruding into the flow and introducing a vertical offset of the fluid and extended streamwise recovery distance.
The current treatment to make denticles embedded is a better representation of realistic shark skin fully covered by denticles.  
Second, we include the neck of the denticle — a slender, cylinder-like structure situated underneath the wide lid-like crown (figure \ref{fig:domain}(a)). Only the bottom 15\% of the denticle that is supposed to be embedded in the underlying stratum spongiosum of the dermis is excluded in the current configuration. On the contrary, most previous studies of shark denticles only consider the crown. For example, the DNS by \cite{BoomsmaF16} ignored the bottom half of the denticle, and the experiments of Domel {\it et al.} discarded the lower 60\% \citep{Domeletal18_low}.
As we will discuss momentarily, the neck region is found to play an important role in determining the drag modulation during flow separation. 

The height of the denticle is chosen to be $\delta_h = 0.0488H$, half of that in the DNS by \cite{BoomsmaF16}.
The staggered denticle array (figure \ref{fig:domain}(b)) is arranged with an offset spacing of $\delta_x=1.032\delta_w$ and $\delta_z=1.142\delta_w$ ($\delta_w$ is the width of the denticle, $z$ denotes the spanwise direction). 
A total of 21 by 50 denticles cover the entire lee side of the bump. 
The 3D model of a representative denticle from {\it Isurus Oxyrinchus} (Shortfin Mako) provided by George Lauder is used~\citep{WenWL14}  (shown in the inset of figure \ref{fig:domain}(a)). It was scanned using micro-CT and made symmetric about the spanwise direction. The model and the spacing are the same as the study by \cite{BoomsmaF16}. 

The simulations are performed using an established solver of the incompressible Navier-Stokes equations on a Cartesian staggered grid \citep{KeatingPBN04}.
\begin{equation}
\dpr{u_i}{x_i}=0; \;\;\dpr{u_i}{t} + {u}_j\dpr{u_i}{x_j}=-\dpr{p}{x_i} + \frac{1}{Re} \frac{\partial ^2 u_i}{\partial x_j^2} + f_i.
\label{eq:NS}
\end{equation}
A second-order Adams-Bashforth scheme is employed for the convective terms while the diffusion terms are discretized using an implicit Crank-Nicolson scheme. The spatial derivatives are computed using a second-order accurate central difference scheme. The Poisson equation is solved with a pseudo-spectral method.
The complex geometries of the embedded objects are represented by unstructured surface mesh with triangular elements. 
The boundary condition on the body surface is applied by an immersed boundary method. At each time step, the forcing term $f_i$ in the right-hand-side of the NS equation~\citep{Scotti06,YuPhD15} is assigned based on the volume-of-fluid (VOF) of each grid cell to satisfy the no-slip boundary condition. 
The solver has shown great success in our previous
studies of surface structures in turbulence including sandgrain roughness (e.g., \citealt{WuPiomelli18}),
bumps (e.g., \citealt{WuPS22}), etc.

The computational domain is $22.4H$ by $2H$ by $3.045H$ in the streamwise ($x$), wall-normal ($y$), and spanwise ($z$) directions, respectively. Spanwise two-point correlations prove that fluctuating velocities are uncorrelated by half of the span both near the wall and in the separated shear layer. 
At the inflow, instantaneous velocity snapshots saved from an {\it a-priori} channel flow at the same $Re_b$~\citep{WuSARB23} are used. The statistics of the inflow data closely match those from previous DNS (\citealt{Lee15}, not shown). 
A convective boundary condition is used at the outflow. %
Both cases use a grid with $N_x \times N_y \times N_z = 2016 \times 415 \times 2016$ grid points. The finest grid spacing is determined by the denticle features, not the viscous scale. The grid is uniform in $z$, and stretched in $x$ and $y$ with local refinement around the bump.
This resolution is comparable to that of~\cite{BoomsmaF16}. Outside of the refined region, the grid is gradually stretched. In wall units, $\Delta x^+<9.0$, $\Delta y^+_{1}<0.2$ and $\Delta z^+<0.25$. In regions distant from the wall, the maximum ratio between the Kolmogorov scale and the grid spacing is below 2 for $x/H \in [1.0, 9.0]$ and remains below 4 elsewhere. This guarantees the resolution of a significant portion of the dissipation spectrum. Grid convergence is verified by the DNS of open channel flows over denticle arrays~\citep{WuSARB23}. The current ($\sim 40^3$ per denticle) and refined ($\sim 70^3$) grids
result in negligible changes in flow statistics and less than 1\% difference in drag. 
The simulations were performed with a constant time step $\Delta t = 2.1 \times 10^{-4} H/U_b$. After a statistically steady state was reached, snapshots were collected every 
$0.5H/U_b$ over $200 H/U_b$ to obtain the statistics. %
Quantities are averaged in time and in the spanwise direction over the fluid domain.%

\begin{figure}
\centering
  \includegraphics[width=12cm]{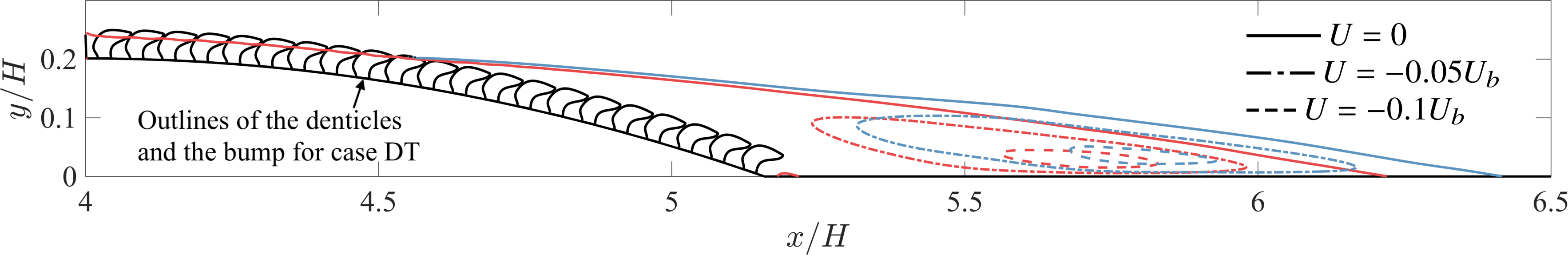}
  \caption{Mean separation region.  Blue, smooth case; red, denticle case. }
  \label{fig:ueq0}
\end{figure}

\section{Results and Discussions}
\subsection{2D mean flow characteristics}
Figure \ref{fig:ueq0} shows the mean separation region in the two cases. Compared to case SM, case DT has a mean separation bubble that is 12\% shorter, mainly due to an earlier reattachment downstream of the bump. 
The separation point, identified as the location of $U=0$ on the smooth bump or at the crest of the denticles, barely changes ($x/H$ = 4.60). 
The reduction of the separation bubble agrees with \cite{Evansetal17}, \cite{Doosttalabetal18} and \cite{Santosetal21}. 
Newly observed is a reverse pore flow (RPF) beneath the denticle crown prior to the onset of massive flow separation (figure \ref{fig:meanreverse}(a)). This pore flow has a relatively small magnitude up to 0.025$U_b$, about half of the friction velocity of the undisturbed channel. Locally, it's peak occurs at $\sim$25\% denticle height from the base and in the spanwise gap between denticle necks. 
Interestingly, the RPF diminishes once the massive separation takes place. We will elaborate on its physical mechanism momentarily. 
It is worth mentioning that the neck region has received limited attention in the context of biomimetic structures of shark denticles and the pore flow has rarely been measured before. \cite{Evansetal17} and~\cite{Doosttalabetal18} addressed it to some extent, albeit without providing data underneath the crown. In our previous study of the same denticle in a turbulent channel \citep{WuSARB23}, the mean pore flow appeared as a weak forward flow up to $0.007U_b$ throughout. Therefore, the RPF here appears to be activated by the APG.

\subsection{Drag reduction and 3D RPF}
The drag produced by the bump is the net outcome of the forces over the wind and lee sides of the bump, the latter of which is expected to be a thrust via the pressure directed upstream and the reversed skin friction by recirculation.  
While the wind-side drag changes little between the two cases, the thrust over the lee side of the bump increases by 230\% in case DT. The total drag reduction is about 4\% (the three-fold increase in thrust remains relatively small compared to the drag over the wind side in current configuration). Here, we focus on the increase of thrust over the lee side since the drag on the wind side is irrelevant to the control functionality of the denticles during the flow separation.

\begin{figure}
\centering
  \includegraphics[width=10cm]{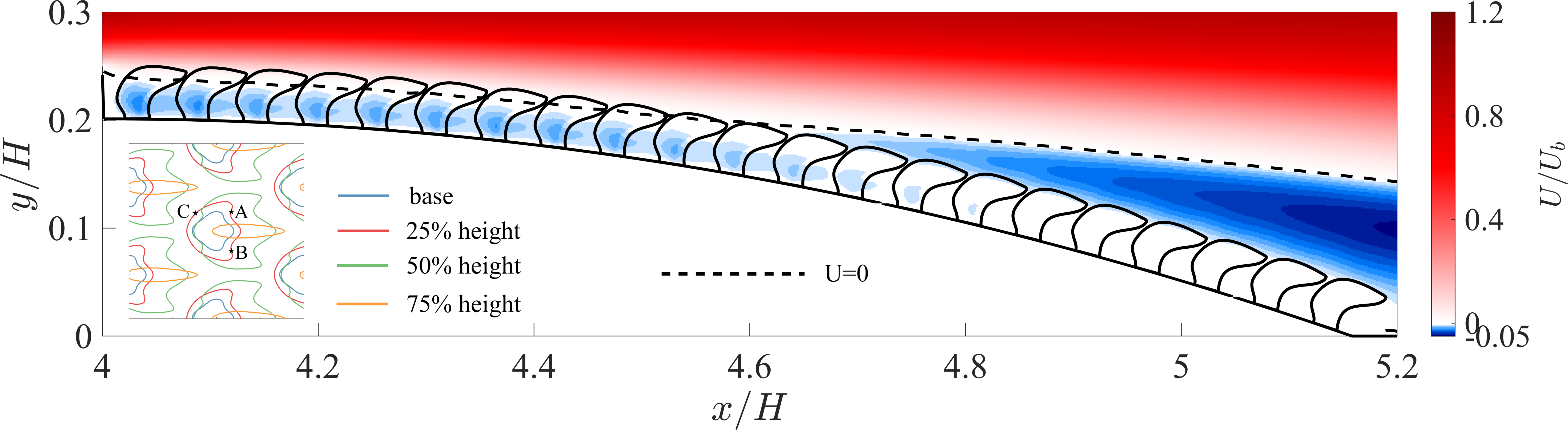}
  \includegraphics[width=10cm]{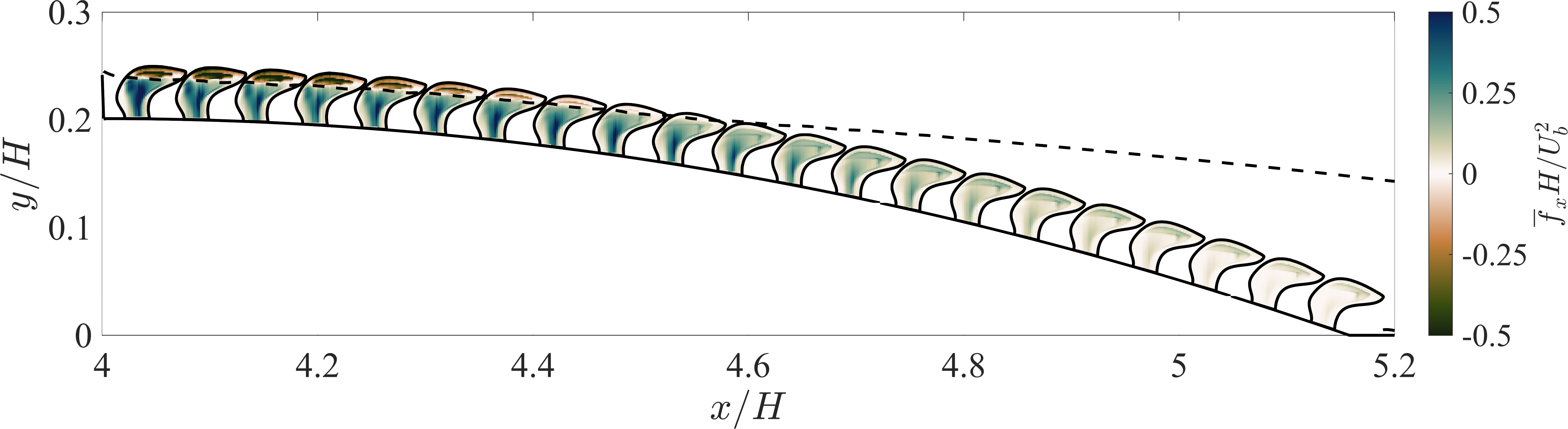}
  \includegraphics[width=10cm]{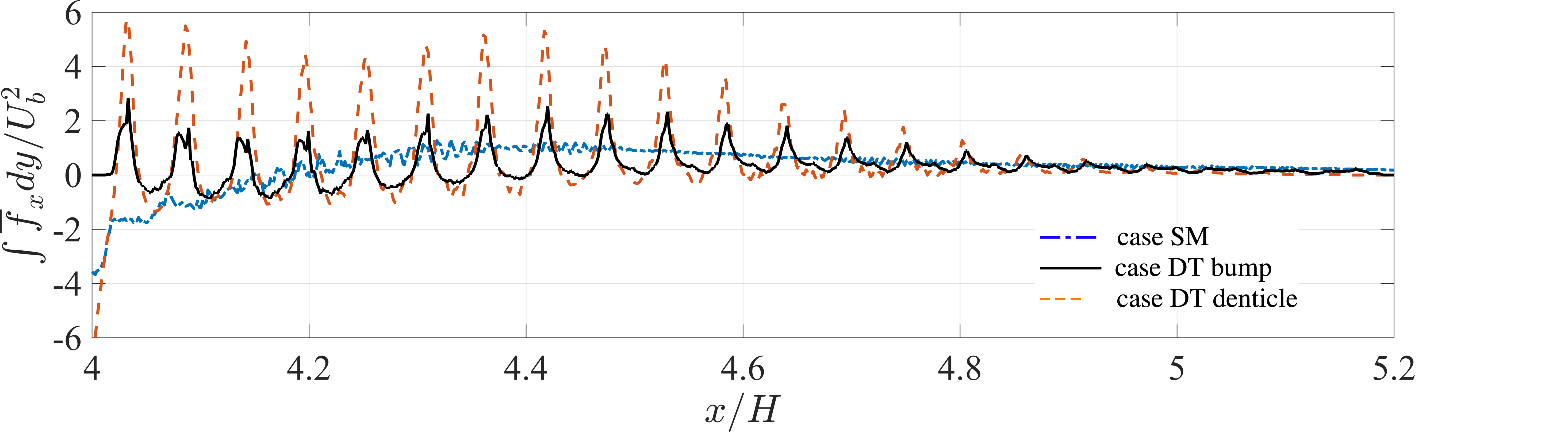}  
  \caption{(a) Mean streamwise velocity over the denticle bump. The inset shows the outlines of a denticle and its neighbors at selected heights from the base. (b) Mean streamwise force exerted by the denticles to the fluid
  (c) Profiles of total force integrated in the wall-normal direction.}
  \label{fig:meanreverse}
\end{figure}

To identify the thrust-generation mechanism, we visualize the distribution of $\bar{f}_x$ in (\ref{eq:NS}) (total force including both pressure and viscous forces). Negative $\bar{f}_x$ indicates the solid is impeding the fluid's motion, thereby producing drag. Conversely, a positive $
\bar{f}_x$ indicates a thrust. The distribution of $\bar{f}_x$ over the denticles is exhibited in figure \ref{fig:meanreverse}(b). Note that the surface of the indented bump that is not covered by the denticle base also contributes to the total force yet is not shown in this contour. It can be seen that the denticles are producing thrust by their necks. 
The ridges on the top of the crowns generate drag up to the mean separation point. The alignment between the drag/thrust boundary and the $U$=0 line indicates that the drag is due to friction.
The spanwise spacing of the ridges in wall units is $s^+=3.6$, \textit{i.e.}, below the optimal riblet spacing for drag reduction ($\sim 16$ viscous lengths) \citep{García-MayoralJ11}.
This, along with the observed drag increase in previous studies of denticle crowns in ZPG flows \citep{BoomsmaF16, WuSARB23}, indicates the complex geometry of the crown may jeopardize the possible riblet drag-reduction dynamics, and highlight the configuration-sensitive nature of drag reduction by denticle crowns.

The most significant thrust over the lee side coincides with the prominent RPF, occurring shortly downstream of the bump crest. For each row, local peak thrust occurs near the streamwise center of the neck and approximately 25\% of the denticle height from the bump. This specific location corresponds to the widest spawise section of the neck, featuring backward-facing bulges on the spanwise sides of the neck (points A and B in the inset of 25\% height in figure \ref{fig:meanreverse}(a)). 
Integrating $\bar{f}_x$ in the vertical direction, the local drag/thrust profile is obtained and shown in figure \ref{fig:meanreverse}(c). For case DT, the contribution of the denticles and the bump are separated for discussion. Compared to the thrust over the smooth bump in case SM, these localized thrusts due to the denticles, in total, are two to five times larger in magnitude. Over the entire lee side, the denticle necks contribute to 
38\% of the total thrust, yet only occupy 13\% of the surface area at the base. In particular, the peak thrust produced by the bump surface aligns with that generated by the denticle neck. 
Therefore, the thrust is strongly correlated to the necks of the denticle and the acceleration of the RPF between them.

\begin{figure}
\centering
   \includegraphics[width=0.67\textwidth]{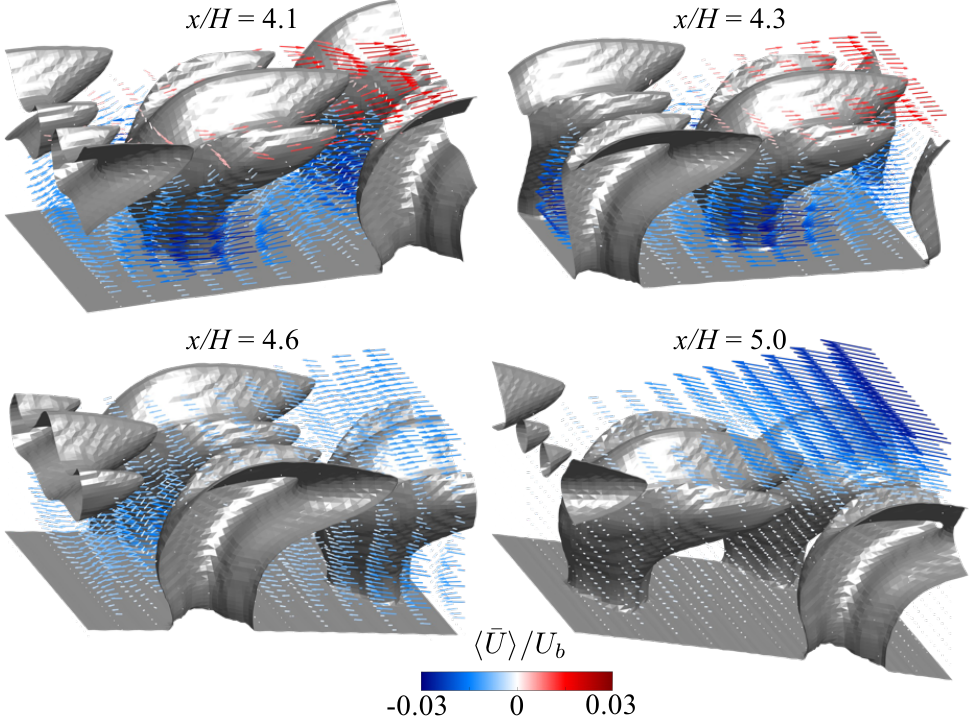}
  \caption{Temporal- and ensemble-averaged velocity vector at selected locations, colored by the streamwise component.}
  \label{fig:ensembleavg}
\end{figure}

We named this process a `channeling effect' of the denticle neck. Its 3D features are extracted by 3D temporal averaging followed by ensemble averaging between the 50 denticles along the span (denoted by operator $\langle \bar{.}\rangle$). Figure \ref{fig:ensembleavg} shows the velocity vector around the denticle at selected streamwise locations, colored by $\langle \bar{u}\rangle$. The forward flow above the crown at $x/H=4.1$ and 4.3 and the prominent RPF in the cavity region are evident. The vector fields confirm that the RPF is accelerated by the narrow gap at the widest section of the necks, 
forming a jet that strikes the rear of upstream denticles. The staggered arrangement of denticle necks in the wall-parallel plane enables the channeling and acceleration of pore flow (and corresponding thrust generation). 
No flow separation is observed over the mildly curved front of the neck (point C in the inset of figure \ref{fig:meanreverse}(a)). The thrust is generated by the friction when the RPF travels around the denticle neck, as well as the stagnation pressure when it impinges on the backward-facing side bulges. 
At $x/H=4.6$ near the mean separation point, fluid is entering the cavity region via the slits between streamwise consecutive denticles. The RPF at this location is less intense than the ones upstream. 
By $x/H=5.0$, the RPF and the penetrating flow are very weak and the channeling effect is negligible.

\subsection{Mechanisms of the RPF}
\begin{figure}
\centering
  \includegraphics[width=0.85\linewidth]{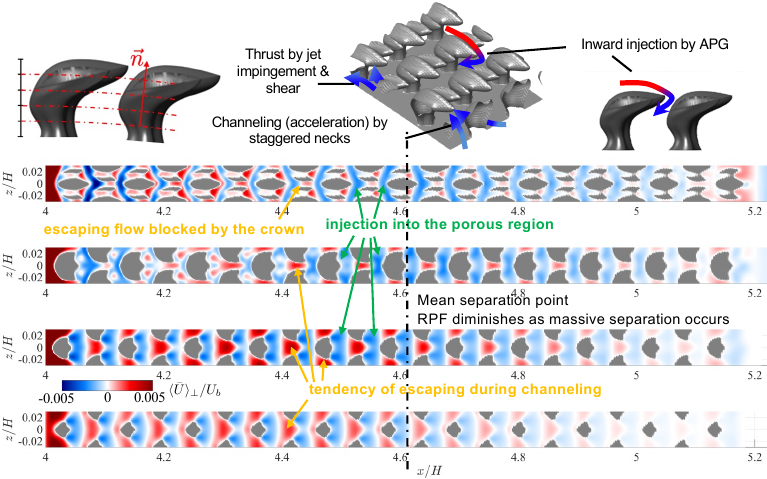}
  \caption{Temporal- and ensemble-averaged velocity in the bump-normal direction. From bottom to top, planes at 20\%, 40\%, 60\% and 80\% denticle height, respectively (refer to upper left inset). The injection of fluid through the backward-upward-facing slit between denticle rows are demonstrated in the middle and upper right insets.}
  \label{fig:plane}
\end{figure}

Understanding how the RPF is initiated and sustained will shed light on utilizing this mechanism to reduce drag in engineering applications. Figure \ref{fig:plane} shows the bump-normal velocity at several virtual planes that are parallel to the bump at selected distances. Notably, negative normal flux is correlated with the slits between denticle rows in the streamwise direction. This inward injection effectively penetrates deep into the cavity region and still evident at 20\% denticle height. 
Referring to the geometry of the denticles and the pattern of the denticle array (figure \ref{fig:domain} and insets of figure \ref{fig:plane}), the slit is formed by the inclined bottom surface of a denticle and the rounded fronts of its downstream neighbors, shaping as an backward-upward-facing slit (denoted as BUFS). 
Meanwhile, a positive normal velocity, which indicates the tendency of pore flow to escape the cavity region, is observed between the necks of the denticles in the spanwise direction. It is strongest at 0.4$\delta_h$ and becomes weaker above it. By $0.8\delta_h$, where the crown reaches it maximum spanwise width, this outward flux has reduced significantly in the spanwise gaps. Some leakage can be observed downstream of the trailing edges of the crown but is not comparable to the dominant inward flux.

\begin{figure}
\centering
  \includegraphics[width=8cm]{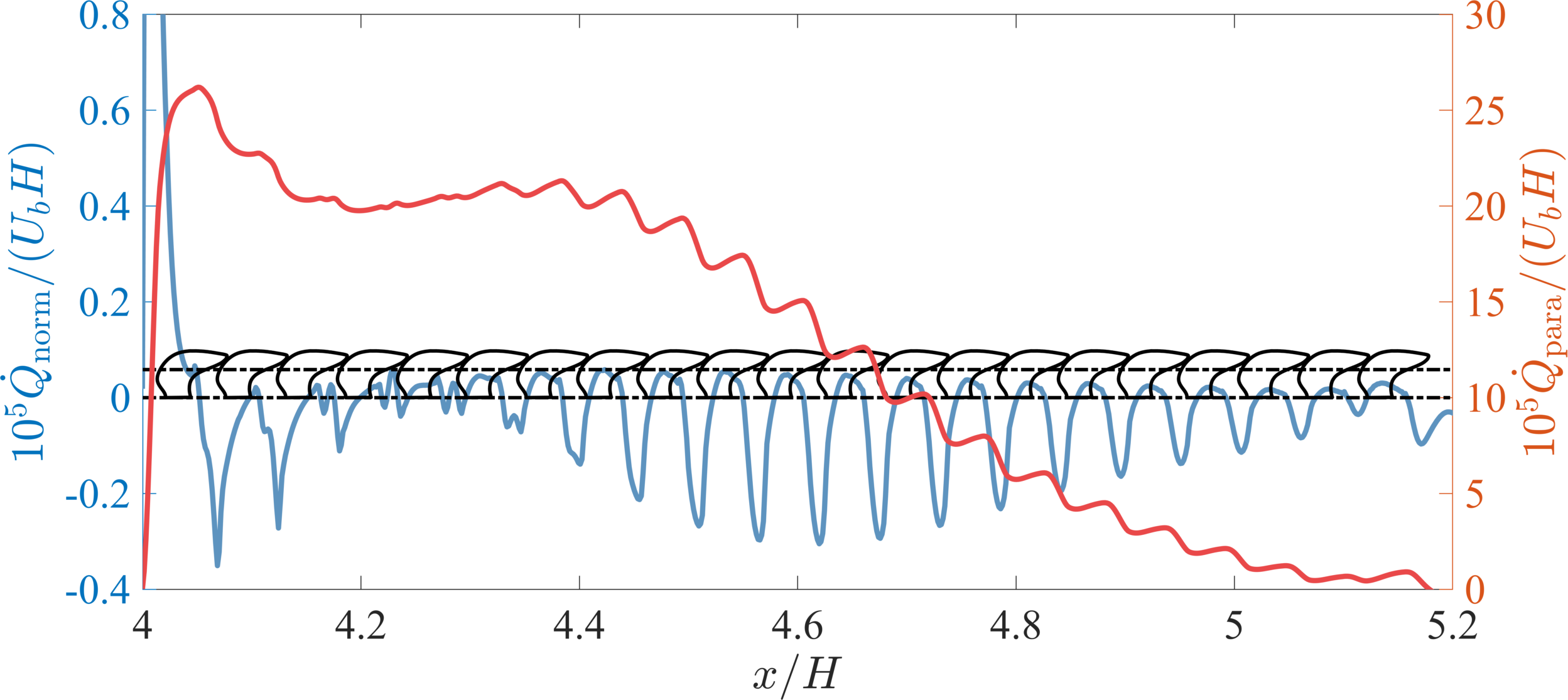}
  \caption{Volume flux in the: left axis with blue line, bump-normal, and right axis for red line, bump-normal directions. A dash-dotted line at 60\% denticle height shows where the bump-normal flux is calculated and the bump-parallel flux is integrated upto. The outline of the denticle array is shown by the solid black line as reference.}
  \label{fig:flux}
\end{figure}

The patterns of the normal flux indicate the following physical process of the formation and sustentation of the RPF. The flow above the denticle enters the underneath cavity region through the BUFS. Then, when the RPF gets accelerated by the channelling effect, it tends to moves upward towards the spanwise gaps above. 
The closed packing of the denticles and the wide crowns narrow the gap to restrict the outward flux. 
Among these features, the directional preference of BUFS appears to be critical as it is where the RPF originates. 
Its backward-upward inclination permits fluid only to enter in reverse. Recall that the pore flow is negligible in a ZPG channel~\citep{WuSARB23}. Therefore, this directional preference of the BUFS enables anisotropic permeability of the cavity region. 

As the essential role of the BUFS is demonstrated, the factor that determines the inward flux becomes the core question to explore. 
While the injection and leakage phenomena are qualitatively robust over the lee side of the bump, the streamwise variations of their magnitudes is evident (refer to figures \ref{fig:meanreverse}(a) and \ref{fig:plane}). It suggests the injection through the BUFS, the RPF, and assoticated drag-reduction are activated/determined by specific flow conditions. 
A few previous studies noticed the pore flow and proposed various assumptions regarding their kinematics. 
\cite{BechertHR85} assumed that low instantaneous local pressure created by near-wall streaks ejects fluid out
through the BUFS (opposite to the inward injection that we observe), and as a result, compensates
the streaks and reduces drag.
\cite{Evansetal17}, in their study with divergent pillars, assumed that the contraction - of forward pore flow - between the pillar and stagnation by the top creates local pressure variations and modulate the turbulent structures in the buffer layer. Our data shows that their forward flow assumption is questionable. 
Lang and co-workers~\citep{Langetal11,Santosetal21} hypothesized that a local cavity recirculation is formed when a reversed flow bristles the denticle. Our results show that the inward flow peaks near the separation point where the flow is near stationary and diminishes in the recirculation region where the reversed flow is strong, contradicting the reversed-flow activation proposed by them.

Figure \ref{fig:flux} shows the momentum fluxes through the cavity region. Both the local bump-normal flux (blue, left axis) and the accumulated bump-parallel flux (red, right axis) are shown. The normal flux, measured across the plane parallel to the bump at 60\% of the denticle height, exhibits negative peaks in the locations of BUFS. Meanwhile, positive normal flux occurs around the necks. These observations further solidify the physical picture of the formation (through the BUFS) and sustentation (leakage prevention by wide crown) of the RPF as discussed above. 
The net penetrating flux is negative in most of the regions except around $x/H=4.3$ where significant leakage is observed, and at $x/H=4.0$ where the RPF reaches the backward-facing indented wall at the beginning of the denticle array. Once the flow separates around $x/H=4.6$, the inward flux decreases despite that the mean reversed flow over the crown becomes stronger. When the flow detaches, the strong recirculation region usually has a nearly ZPG as observed extensively in aerodynamics as a plateau in the surface pressure \citep{Patrick87}. The mean streamwise pressure gradient in case DT agrees with this expected distribution: the streamwise pressure gradient is decreased by 60\% by $x/H=4.8$ and 85\% by $x/H=5.0$, compared to it at the separation point (not shown). 
Therefore, the inward flux and RPF are not associated with the direction of the flow over the denticle crown. Rather, they are determined by the APG. 

Therefore, the inward flux through the BUFS is driven by the APG: it is initiated near the separation point; as the RPF moves upstream on the lee side, it 
is continuously fed by successive BUFS jets, leading to the increase of the accumulated bump-parallel flux. 
It explains the acceleration of RPF towards the bump crest shown in figure \ref{fig:meanreverse}. 
As the RPF approaches the bump crest, the leakage becomes significant. 
Due to mass conservation, the accumulated flux must exit the cavity region at some point. A weakening APG approaching the start of the geometric expansion and/or an upstream FPG could affect the release of the RPF. The current configuration is not able to justify this process. Future work is warranted.

\section{Discussion and concluding remarks}
The present study focuses on the effects of shark denticles during flow separation, especially the cavity region underneath the denticle crown. Such region has been noticed in early studies of shark skins \citep{BechertHR85} yet is much less explored in recent investigations. Our result quantifies the flux penetrating and within the cavity region with detailed data for the complete shark scale patterns. It enables the justification and refinement of several inaccurate hypothetical mechanisms proposed in the literature. 
The key findings include the APG activation of the flux into the cavity region and the channeling effect of the reversed pore flow (RPF) that enables the substrate thrust. 

We have identified several geometrical features of the denticle that influence the thrust generation by the RPF: 1) the backward-upward facing slits (BUFS), 2) the staggered arrangement, 3) the crown width, and 4) the non-circular neck. The BUFS facilitates a penetrating flux that is only enabled under APG, avoiding drag penalty by a forward pore flow under FPG/ZPG. The staggered arrangement of denticle necks enables the channeling and acceleration of the RPF. A wide crown narrows the spanwise gaps between adjacent denticles when closely packed. This restriction minimizes outward leakage as the neck channels the RPF, thus sustaining the thrust generation. Finally, the denticle neck features a cross-section possessing wide side bulges that enhance RPF impaction.
It appears that sharks may have developed their denticles into a complex geometry with multiple hydrodynamic functions - more than just the ridges over the crown - through their long evolution history.
The activation and sustain of the RPF by APG indicates that its thrust production/drag reduction mechanism is best suited for situations experiencing mild APGs and no flow separation.
Note that an APG is often inevitable due to diverging surface curvature or maneuvers. The RPF-thrust generation leverages and transforms an APG into a favorable outcome. This offers an exciting new strategy for drag reduction. 

Numerous studies on porous media flows have investigated fluid flows across cylinders and characterized their dynamics. In the current configuration, the denticle necks form a porosity around 0.85 at $0.25\delta_h$. According to~\cite{KhalifaPT20}, among others, the Stokes flow between staggered cylinder arrays at this porosity level is Darcian for $Re_D = U_\mathrm{pore} D/\nu <1.15$ ($D$ is the diameter of the neck, $U_\mathrm{pore}$ the mean pore flow velocity).  Pore scale flow inertial effects becomes important for $Re_D>$1.78, and vortex shedding starts at $Re_D = 31$. $Re_D$ in our case DT is 1.56, thus the flow is in the transitional condition between the Darcy and Forchheimer regimes. 
The permeability can be estimated using the %
Darcy-Forchheimer Law, 
\begin{equation}
-Re_D \frac{dP}{dx} = \frac{1}{K_F} + Re_D \frac{\alpha}{\sqrt{K_F} }. 
\label{eq:DFeq}
\end{equation}
Here, $K_F$ is the Forchheimer permeability and $\alpha$ the Forchheimer coefficient. 
$K_F = 0.19 D^2$ (using $\alpha = 0.121$~\cite{KhalifaPT20}) is obtained using the present data. These values qualitatively agree with the ones found by~\cite{KhalifaPT20} for staggered cylinder arrays. 
{\color{black} For the denticles in a channel~\citep{WuSARB23}, on contrary, 
$K = 0.005 D^2$, indicating the denticles are not permeable without the APG. 

We have to stress that the 4\% drag reduction observed should not be interpreted as the actual efficiency outcome for sharks' locomotion. The configuration differs from a real shark skin and swim conditions.
The reason for the negligible change in the onset of separation between the two cases may be the APG that is strong along the expanding lee side of the bump. The average expansion rate of the cross-section, for example, is 0.15 in \cite{Evansetal17} and \cite{Doosttalabetal18}, while ours is 0.19. 
Sharks may have a strategy to maintain a mild curvature and APG thus a prolonged utilization of the RPF-thrust generation and possible separation reduction.  
The current findings do not rule out the possible effect of denticle bristling.
Rather, it highlights what occurs before the massive flow separation. 

\section*{Acknowledgements}
The authors acknowledge the NSF EPSCoR RII Track-4 Program (grant OIA-2131942) and TACC Stampede2. BS acknowledges the support from the NSF GRFP (grant 2235036). 

\section*{Declaration of interest}
The authors declare no conflict of interest.


\bibliographystyle{./jfm}
\bibliography{myref}

\begin{thebibliography}{25}
\expandafter\ifx\csname natexlab\endcsname\relax\def\natexlab#1{#1}\fi
\def\au#1{#1} \def\ed#1{#1} \def\yr#1{#1}\def\at#1{#1}\def\jt#1{\textit{#1}}
  \def\bt#1{#1}\def\bvol#1{\textbf{#1}} \def\vol#1{#1} \def\pg#1{#1}
  \def\publ#1{#1}\def\arxiv#1{#1}\def\org#1{#1}\def\st#1{\textit{#1}}

\bibitem[Arunvinthan {\em et~al.\/}(2021)Arunvinthan, Raatan, {Nadaraja
  Pillai}, Pasha, Rahman \& Juhany]{Arunvinthanetal21}
{\sc \au{Arunvinthan, S.}, \au{Raatan, V.~S.}, \au{{Nadaraja Pillai}, S.},
  \au{Pasha, A.~A.}, \au{Rahman, M.~M.} \& \au{Juhany, K.~A.}} \yr{2021}
  \at{Aerodynamic characteristics of shark scale-based vortex generators upon
  symmetrical airfoil}.  \jt{Energies}  \bvol{14},  \pg{1808}.

\bibitem[Bechert {\em et~al.\/}(1985)Bechert, Hoppe \& Reif]{BechertHR85}
{\sc \au{Bechert, D.}, \au{Hoppe, G.} \& \au{Reif, W.~E.}} \yr{1985} On the
  drag reduction of the shark skin.  \bt{In {\em 23rd Aerospace Sci.
  Meeting\/}},  \pg{pp. 1--18}.

\bibitem[Boomsma \& Sotiropoulos(2016)]{BoomsmaF16}
{\sc \au{Boomsma, A.} \& \au{Sotiropoulos, F.}} \yr{2016}  \at{Direct numerical
  simulation of sharkskin denticles in turbulent channel flow}.
  \jt{Phys.~Fluids}  \bvol{28},  \pg{035106}.

\bibitem[Chen {\em et~al.\/}(2023)Chen, Du, Li, Lv \& Duan]{Chenetal23}
{\sc \au{Chen, G.-Q.}, \au{Du, Z.-Z.}, \au{Li, H.-Y.}, \au{Lv, P.-Y.} \&
  \au{Duan, H.-L.}} \yr{2023}  \at{{On the drag reduction of an inclined wing
  via microstructures with the immersed boundary-lattice Boltzmann flux
  solver}}.  \jt{Phys.~Fluids}  \bvol{35},  \pg{087105}.

\bibitem[Domel {\em et~al.\/}(2018)Domel, Saadat, C.Weaver, Haj-Hariri,
  Bertoldi1 \& Lauder]{Domeletal18_low}
{\sc \au{Domel, A.~G.}, \au{Saadat, M.}, \au{C.Weaver, J.}, \au{Haj-Hariri,
  H.}, \au{Bertoldi1, K.} \& \au{Lauder, G.~V.}} \yr{2018}  \at{Shark
  skin-inspired designs that improve aerodynamic performance}.
  \jt{J.~R.~Soc.~Interface.}  \bvol{15},  \pg{20170828}.

\bibitem[Doosttalab {\em et~al.\/}(2018)Doosttalab, Dharmarathne, Evans, Hamed,
  Gorumlu, Aksak, Chamorro, Tutkun \& Castillo]{Doosttalabetal18}
{\sc \au{Doosttalab, A.}, \au{Dharmarathne, S.}, \au{Evans, H.~B.}, \au{Hamed,
  A.~M.}, \au{Gorumlu, S.}, \au{Aksak, B.}, \au{Chamorro, L.~P.}, \au{Tutkun,
  M.} \& \au{Castillo, L.}} \yr{2018}  \at{{Flow modulation by a mushroom-like
  coating around the separation region of a wind-turbine airfoil section}}.
  \jt{J.~Renew.~Sustain.~Energy}  \bvol{10},  \pg{043305}.

\bibitem[Evans {\em et~al.\/}(2018)Evans, Hamed, Gorumlu, Doosttalab, Aksak,
  Chamorro \& Castillo]{Evansetal17}
{\sc \au{Evans, H.~B.}, \au{Hamed, A.~M.}, \au{Gorumlu, S.}, \au{Doosttalab,
  A.}, \au{Aksak, B.}, \au{Chamorro, L.~P.} \& \au{Castillo, L.}} \yr{2018}
  \at{Engineered bio-inspired coating for passive flow control}.
  \jt{Proc.~Natl.~Acad.~Sci.~U.S.A.}  \bvol{115},  \pg{1210--1214}.

\bibitem[{García-Mayoral} \& Jim\'{e}nez(2011)]{García-MayoralJ11}
{\sc \au{{García-Mayoral}, R.} \& \au{Jim\'{e}nez, J.}} \yr{2011}  \at{Drag
  reduction by riblets}.  \jt{Philos.~Trans.~Royal Soc.~A}  \bvol{369},
  \pg{1412--1427}.

\bibitem[Guo {\em et~al.\/}(2021)Guo, Zhang, Yasuda, Yang, Galipon \&
  Rival]{Guoetal21}
{\sc \au{Guo, P.}, \au{Zhang, K.}, \au{Yasuda, Y.}, \au{Yang, W.}, \au{Galipon,
  J.} \& \au{Rival, D.~E.}} \yr{2021}  \at{On the influence of biomimetic shark
  skin in dynamic flow separation}.  \jt{Bioinspir.~Biomim.}  \bvol{16},
  \pg{034001}.

\bibitem[Keating {\em et~al.\/}(2004)Keating, Piomelli, Bremhorst \&
  Ne\v{s}i\'{c}]{KeatingPBN04}
{\sc \au{Keating, A.}, \au{Piomelli, U.}, \au{Bremhorst, K.} \&
  \au{Ne\v{s}i\'{c}, S.}} \yr{2004}  \at{Large-eddy simulation of heat transfer
  downstream of a backward-facing step}.  \jt{J.\ Turbul.}  \bvol{5},  \pg{20}.

\bibitem[Khalifa {\em et~al.\/}(2020)Khalifa, Pocher \& Tilton]{KhalifaPT20}
{\sc \au{Khalifa, Z.}, \au{Pocher, L.} \& \au{Tilton, N.}} \yr{2020}
  \at{Regimes of flow through cylinder arrays subject to steady pressure
  gradients}.  \jt{International Journal of Heat and Mass Transfer}
  \bvol{159},  \pg{120072}.

\bibitem[Lang {\em et~al.\/}(2011)Lang, Motta, Habegger, Hueter \&
  Afroz]{Langetal11}
{\sc \au{Lang, A.}, \au{Motta, P.}, \au{Habegger, M.~L.}, \au{Hueter, R.} \&
  \au{Afroz, F.}} \yr{2011}  \at{Shark skin separation control mechanisms}.
  \jt{Mar.~Technol.~Soc.~J.}  \bvol{45},  \pg{208--215}.

\bibitem[Lang {\em et~al.\/}(2008)Lang, Motta, Hidalgo \& Westcott]{Lang_2008}
{\sc \au{Lang, A.~W.}, \au{Motta, P.}, \au{Hidalgo, P.} \& \au{Westcott, M.}}
  \yr{2008}  \at{Bristled shark skin: a microgeometry for boundary layer
  control?}  \jt{Bioinspir.~Biomim.}  \bvol{3},  \pg{046005}.

\bibitem[Lee \& Moser(2015)]{Lee15}
{\sc \au{Lee, M.} \& \au{Moser, R.~D.}} \yr{2015}  \at{Direct numerical
  simulation of turbulent channel flow up to ${R}e_{\tau} \approx 5,200$}.
  \jt{J.~Fluid~Mech.}  \bvol{774},  \pg{395--415}.

\bibitem[Lloyd {\em et~al.\/}(2023)Lloyd, Mittal, Dutta, Dorrell, Peakall,
  Keevil \& Burns]{Lloydetal23}
{\sc \au{Lloyd, C.~J.}, \au{Mittal, K.}, \au{Dutta, S.}, \au{Dorrell, R.~M.},
  \au{Peakall, J.}, \au{Keevil, G.~M.} \& \au{Burns, A.~D.}} \yr{2023}
  \at{Multi-fidelity modelling of shark skin denticle flows: insights into drag
  generation mechanisms}.  \jt{Royal Soc.~Open Sci.}  \bvol{10},  \pg{220684}.

\bibitem[Oeffner \& Lauder(2012)]{OeffnerL12}
{\sc \au{Oeffner, J.} \& \au{Lauder, G.~V.}} \yr{2012}  \at{{The hydrodynamic
  function of shark skin and two biomimetic applications}}.  \jt{J.~Exp.~Biol.}
   \bvol{215},  \pg{785--795}.

\bibitem[Patrick(1987)]{Patrick87}
{\sc \au{Patrick, W.~P.}} \yr{1987} Flowfield measurements in a separated and
  reattached flat plate turbulent boundary layer.  \bt{In {\em
  NASA~Tech.~Rep.~\/}},  \pg{p. 4052}.

\bibitem[Reif(1985)]{Reif85}
{\sc \au{Reif, W.~.E.}} \yr{1985}  \bt{Squamation and ecology of sharks}. {\em
  Tech. Rep.\/}.  \org{Courier Forschungsinstitut Senckenberg}, Frankfurt,
  Germany.

\bibitem[Santos {\em et~al.\/}(2021)Santos, Lang, Wahidi, Bonacci, Gautam,
  Devey \& Parsons]{Santosetal21}
{\sc \au{Santos, L.~M.}, \au{Lang, A.}, \au{Wahidi, R.}, \au{Bonacci, A.},
  \au{Gautam, S.}, \au{Devey, S.} \& \au{Parsons, J.}} \yr{2021}  \at{Passive
  separation control of shortfin mako shark skin in a turbulent boundary
  layer}.  \jt{Exp.~Therm.~Fluid Sci.}  \bvol{128},  \pg{110433}.

\bibitem[Savino {\em et~al.\/}(2023)Savino, Patel \& Wu]{WuPS22}
{\sc \au{Savino, B.}, \au{Patel, D.} \& \au{Wu, W.}} \yr{2023} Reynolds-number
  dependence of separating flow over a bump in spanwise rotating channel flows.
   \bt{In {\em Direct and Large-Eddy Simulation XIII\/}},  \pg{pp. 41--46}.

\bibitem[Scotti(2006)]{Scotti06}
{\sc \au{Scotti, A.}} \yr{2006}  \at{Direct numerical simulation of turbulent
  channel flows with boundary roughened with virtual sandpaper}.  \jt{Phys.\
  Fluids}  \bvol{18},  \pg{031701}.

\bibitem[Wen {\em et~al.\/}(2014)Wen, Weaver \& Lauder]{WenWL14}
{\sc \au{Wen, L.}, \au{Weaver, J.~C.} \& \au{Lauder, G.~V.}} \yr{2014}
  \at{Biomimetic shark skin: design, fabrication and hydrodynamic function}.
  \jt{J.~Exp.~Biol.}  \bvol{217},  \pg{1656--1666}.

\bibitem[Wu \& Piomelli(2018)]{WuPiomelli18}
{\sc \au{Wu, {W}.} \& \au{Piomelli, U.}} \yr{2018}  \at{Effects of surface
  roughness on a separating turbulent boundary layer}.  \jt{J.~Fluid~Mech.}
  \bvol{841},  \pg{552--580}.

\bibitem[Wu \& Savino(2023)]{WuSARB23}
{\sc \au{Wu, W.} \& \au{Savino, B.}} \yr{2023} Dynamics of pore flow between
  shark dermal denticles.  \bt{In {\em Annual Research Briefs\/}},  \pg{pp.
  195--206}.  \publ{Center for Turbulence Research, Stanford University}.

\bibitem[Yuan(2015)]{YuPhD15}
{\sc \au{Yuan, J.}} \yr{2015}  \at{Numerical simulations of rough-wall
  turbulent boundary layers}. PhD thesis, Queen’s Uiverity.

\end{thebibliography}

\end{document}